# Route to stabilize cubic gauche polynitrogen to ambient conditions via surface-saturation by hydrogen

Guo Chen,[1,2,‡] Caoping Niu,[1,2,‡] Wenming Xia,[1,2] Jie Zhang,[1] Zhi Zeng,[1,2] and Xianlong Wang[1,2,*]

[1]*Key Laboratory of Materials Physics, Institute of Solid State Physics, HFIPS, Chinese Academy of Sciences, Hefei 230031, China*

[2]*University of Science and Technology of China, Hefei 230026, China*

___________

[*]Author to whom all correspondence should be addressed: xlwang@theory.issp.ac.cn

[‡]These authors contributed equally to this work.

# Abstract

Cubic gauche polynitrogen (cg-N) is an attractive high-energy density material. However, high-pressure synthesized cg-N will decompose at low-pressure and cannot exist at ambient conditions. Here, the stabilities of cg-N surfaces with and without saturations at different pressures and temperatures are investigated systematically. Pristine surfaces at 0 GPa are very brittle and will decompose at 300 K, especially (1 1 0) surface will collapse completely just after structural relaxation, whereas the decompositions of surfaces can be suppressed by applying pressure, indicating that surface instability causes the cg-N decomposition at low-pressure. Due to the saturation of dangling bonds and transferring electrons to the surfaces, saturation with H can stabilize surfaces at ambient conditions, while OH saturation cannot because of getting electrons from the surfaces. An acidic environment or surface saturation with less electronegative adsorbates is more favorable for the stability of polymerized nitrogen.

**Keywords:** cg-N, high-energy density material, surface saturation, first-principles method, molecular dynamics.

# 1. Introduction

The triple bond in nitrogen molecule (N≡N with bonding energy of 954 kJ/mol) is one of the strongest chemical bonds of all molecules, and the bonding energy will be decrease and total energy will be increase significantly if the N≡N is transformed into the double bond (N=N with bonding energy of 418 kJ/mol) and the single bond (N−N with bonding energy of 160 kJ/mol) (1). On the other hand, an enormous amount of energy will be released when the N−N and N=N decomposes into the N≡N. Therefore, polymerized nitrogen containing N−N or N=N have been attracted lots of interests for potential applications as high-energy density material (HEDM) (1–41).

To break N≡N and synthesize polymerized nitrogen, extreme synthetic conditions are needed, and high pressure is a popular way (2). Existing works shown that within the low pressure range, nitrogen exist in the form of molecule crystals (2–4). The first polymerized nitrogen with fully single-bonding, named as cubic gauche polynitrogen (cg-N), was theoretically predicted to appear at 50 GPa (4), and its energy density is about five times larger than that of TNT (5). Furthermore, many kinds of polynitrogens (such as chains (6,7), layers (8,9), cage (10), and polynitrogen (11,12) )were theoretically proposed to be thermodynamically stable at high pressure. At 2004, the cg-N was successfully synthesized at the extreme conditions of 110 GPa and 2000 K[13]. Following cg-N, layered polymetric nitrogen (LP-N) (14), black phosphorus nitrogen (BP-N) (15), and hexagonal layered polymeric nitrogen (HLP-N) (16) have been synthesized at the pressures of 120 GPa, 150 GPa, and 180 GPa, respectively. Nevertheless, the synthetic pressure of cg-N is the smallest among these synthesized polymeric all-nitrogen materials.

Importantly, there are no-negative frequencies in the phonon spectra of cg-N at 0 GPa (17), indicating that it potentially can be quenched to the ambient pressure. However, the cg-N became unstable and began to decompose with pressure decreasing to 42 GPa at room temperature or to 25 GPa at temperature of 140 K (13), and the same phenomenon also occurs to other polymeric all-nitrogen materials. For example, LP-N, HLP-N, and BP-N can only be stabilized up to 52 GPa, 66 GPa, and 48 GPa at room temperature, respectively (14–16), which hinder their practical applications as HEDMs. In order to preserve the cg-N to the ambient pressure, at present, the important tasks are to find out its decomposition mechanism at low pressure and to find a way to stabilize it at ambient conditions.

It is generally believed that the decomposition of material will start from the weakest bonds, which may exist on the surfaces containing energy unfavorable dangling bonds (42). A typical example is that reconfigurations are usually observed on the material surfaces because of the instability of free surfaces, such as Au (1 1 1) and Si (1 0 0) (43,44). The weakest bond of cg-N may also exist on the surface. Actually, the stability is improved when the cg-N is placed in the carbon nanotubes indicating that surface stability may strongly affect the stability of cg-N (18). However, so far, the surface stabilities of cg-N are not will illustrated.

In this work, the first-principles method as implemented in the Vienna *ab initio* Approximation Package (VASP) (45) combined with molecular dynamics (MD) method is used to investigate the stabilities of cg-N low-index surfaces at different pressures (0 GPa, 40 GPa, 80 GPa, 100 GPa, 120 GPa) and with different saturated conditions (H, OH, $CH_3$). Our results show that the instabilities of pristine surfaces cause the cg-N decomposition at low pressure. The saturation of OH can't improve the surface stabilities, while all investigated surfaces can be stabilized at ambient conditions by H-saturation due to the electron transform from H to surfaces.

## 2. Results

### 2.1 Stability of pristine surfaces

The crystal structure of cg-N is shown in Fig. S1a, where the local coordination environments of N is similar to that of $NH_3$. Each N atom connects with other three neighbors via N−N single bonds, and two electrons form a lone pair of electrons. The N−N single bonds are connected to each other forming a cubic gauche configuration (4). All sub-surfaces obtained from the miller indices of (1 0 0), (1 1 0), and (1 1 1) are shown in Fig. S1b-h. By exposing different surfaces to vacuum, two (100a and 100b) and four (110a, 110b, 110c, and 110d) kinds of surfaces are constructed based on the miller indices of (1 0 0) and (1 1 0), respectively, and only one surface (111a) is obtained in the case of (1 1 1) miller indices. The corresponding free surfaces constructed are shown in Fig. S2a.

The configurations of pristine surfaces after structural relaxation at 0 GPa are shown in Fig. 1a. The 100a and 111a surfaces are stable, keeping their original structures, while linear $N_3$ groups (similar to the azide ion) are formed and adsorbed on 110c surface giving rise to a stable reconfigured 110c (r-110c) surface. After releasing one nitrogen molecule, the 100b surface transfer to the stable 100a surface. Similar phenomenon occurs in the 110a surface, and it will transfer to the r-110c surface after releasing two nitrogen molecules. Interestingly, the 110b surface breaks down layer by layer via releasing nitrogen molecules and decomposes completely under the processes of structure relaxation. The 110d surface is also dissociative completely as well, breaking down into zigzag nitrogen chains and nitrogen molecules. Therefore, the 110b and 110d surfaces are the weakest among that of all investigated sub-surfaces. Following, we investigate the stability of 110b and 110d surfaces under high pressure.

To apply hydrostatic pressure on the surfaces, helium gas is filled in the vacuum zone. The relaxed structures of 110d surfaces at high pressure are shown in Fig. 1b, we can find that at the pressures of 100 GPa and 120 GPa closed to the conditions of experimentally synthesized cg-N, the 110d surface is stable. Furthermore, the structure is maintained with pressure decreasing to 60 GPa, and slight surface-decomposition appears at 40 GPa and 20 GPa, and totally structural decomposition occurs at 0 GPa. As shown in Fig. S3, the stability of 110b surface is also significantly enhanced by applying high pressures larger than 20 GPa. The results consist well with experimental observations, where the cg-N can only be quenched to 25 GPa at low temperature (13). Based on the above results, we can make the following assumption about the process of cg-N decomposition: Since pressure can enhance the stability of surfaces, the surface decompositions will be suppressed at high-pressure condition resulting in a stable cg-N, and surfaces eventually become unstable after depressurization and disintegrate the whole structure.

Furthermore, the stabilities of (100a, 111a, and r-110c) surfaces at 0 GPa and 300 K by MD simulations, The corresponding z-axis mean square displacement (MSD) are shown in Fig. S4. We can find that the MSD of both 100a and 111a surfaces at 300 K increases noticeably, and corresponding N−N networks are broken down from surface after MD simulations (Fig. 1c) indicating that 100a and 111a surfaces cannot exist stably at 300 K. In contrast, the MSD of r-110c surface at 300 K is almost a constant, and its structure can exist stably. Above results show that except r-110c surface, all investigated pristine surfaces are not stable at ambient conditions.

## 2.2 Effects of surface saturations

To stabilize the high-pressure synthesized cg-N to the ambient conditions and promote its practical applications, it is necessary to find a way to stabilize the unstable surfaces (especially 110b and 110d surfaces), and to enhance the finite-temperature stabilities of all investigated surfaces. The high stability of r-110c surface, where the dangling bonds on the surface are saturated with the linear $N_3$ groups, indicates that saturation of the dangling bonds is important for enhancing the stabilities of cg-N surfaces. Therefore, saturations of dangling bonds at 0 GPa with H, OH, and $CH_3$ are investigated, and corresponding saturated surfaces are named as surface-H, surface-OH, and surface-$CH_3$, respectively (Fig. S2).

The stabilities of surfaces with (without) saturations at static condition and at 300 K are summarized in Fig. 2. The OH saturation does not improve the surface stabilities, since the stabilities of 100b, 110d, and 111a surfaces are enhanced but 100a and r-110c surfaces changes from stable to unstable, while the others remains unchanged. Furthermore, the $CH_3$ saturation provides finite stabilizing effect that all surfaces are improved except 100a-$CH_3$ surface, and three surfaces (100b, 110b and 110c) cannot exist stably at 300 K. However, with H saturation, all investigated surfaces are stable at 300 K, indicating that the cg-N can be stabilized to ambient conditions if its surfaces are saturated with hydrogen. Please note that for two stable pristine surfaces (100a and 111a), the H adsorption energies of 100a-H and 111a-H are -0.794 eV and -1.278 eV, respectively, and H saturation is an energy-favorable process.

We further run MD simulations for surface-H at higher temperatures of 500 K, 750 K, 1000 K, and 1250 K for illustrating the maximum temperature that can be stabilized by H saturation, which is important for practical implication as HEDM. The corresponding z-axis MSD at temperatures before and after the structure loses stability are shown in Fig. 3a. We can find that H saturation can improve the stability of all surfaces to at least 750 K and even more than half of the surfaces (100a, 100b, 110a and 110d) can be stabilized to 1000 K. The stabilities of surface-H are summarized in Fig. 3b including that of pristine surfaces used for comparison, indicating that H saturation can enhance the stabilities of cg-N surfaces significantly.

Interestingly, reconfiguration occurring in the pristine 110c surface with the formation of $N_3$ groups (Fig. 1a) is suppressed by H saturation. As shown in Fig. 4, after structural relaxation, the reconfiguration does not occur in the H-saturated 110c. With temperature increasing to 300 K and 500 K, the reconfiguration is triggered by temperature and occurs in half of 110c-H surface, and it further extends to the whole surface at 750 K. Note that, the reconfiguration of 110c surface cannot be suppressed by the $CH_3$ saturation, and $CH_3$ will be decomposed by the reconfiguration (Fig. S5). This phenomenon makes us more convinced that H saturation is a better way to stabilize the cg-N surfaces.

## 2.3 Charge transfer analysis

Based on above discussions, we can find that although the dangling bonds are saturated in both surface-OH and surface-H, their stabilities show large difference. To understand this phenomenon, we further investigate the charger transfer between saturated groups and surfaces relying on the Bader charge analysis (Supplementary table S1), and the average number of electrons transfer from saturated groups into nitrogen atoms are shown in the brackets on the y-axis tag in Fig. 2. Positive (negative) values mean electrons transfer from saturated groups to surfaces (from surfaces to saturated groups). For surface-H, surface-$CH_3$ and surface-OH, the charge transfer is 1.39 e, 1.19 e, and -0.28 e, respectively, and the stability of surfaces increases with the number of electrons on the surfaces (surface-OH < surface-CH$_3$ < surface-H). The electronegativity of OH (3.49) (46) is larger than that of N (3.04) (47), and electrons would transfer from N to OH. However, opposite process occurs when hydrogen atoms are adsorbed on the surface, because the electronegativity of hydrogen (2.20) (50) is notably smaller.

To further rationalize the charge transfer behaviors of surface-OH and surface-H, the electron density difference of 111a-H and 111a-OH surfaces are calculated as shown in Fig. 5, which confirm charge transfers from H to N in 111a-H surface and N to OH in 111a-OH surface. The results clearly show that electrons transfer to nitrogen is beneficial to enhance the surface stabilities, while the opposite leads to decrease the stability. Therefore, although the dangling bonds are saturated in surface-OH, the OH saturation takes electrons from the surfaces and cannot promote their stabilities. However, H adsorption not only saturate the dangling bonds but also transfer electrons to the surfaces, giving rise to a very high stability of cg-N surfaces.

## 3. Conclusion

In summary, based on the first-principles and molecular dynamics method, stabilities of cg-N surfaces are systematically investigated, and effects of pressure, temperature, and saturations with H, OH, and CH$_3$ on the surface stabilities are presented. For pristine cases, only the reconfigured 110c surfaces with linear N$_3$ groups adsorption can exist at 300 K, and other surfaces are very brittle, especially the 110b and 110d surfaces will collapse totally just after structural relaxation. However, the decompositions of surfaces will be suppressed at high-pressure, indicating that surface instability causes the cg-N instability at low pressure. Interestingly, by H saturation, all of investigated surfaces can be stabilized not only at ambient conditions but also at 750 K, since H atoms can saturate dangling bonds combined with transferring electrons to the surfaces. OH saturation cannot improve the stabilities of surfaces, because the it gets electrons from the surfaces. The results show that an acidic environment is very helpful to improve the stabilities of cg-N and its relatives. Furthermore, materials, which can more conducive to transferring electrons to neighbors, such as graphene and carbon nanotubes with as small a band gap as possible, can be used as assistant to improve the stabilities of polymeric nitrogen.

## 4. Materials and Methods

*4.1 Density functional theory*

The first-principles calculations are carried out based on the density functional theory (DFT) method with a plane-wave basis set as implemented in the Vienna *ab initio* Approximation Package (VASP) (45). The exchange-correlation based on the generalized gradient approximation (GGA) (48), parametrized by the Perdew-Burke-Ernzerhof (PBE) functional (49) and projector-augmented-augmented plane-wave (PAW) potentials are used. The energy cutoff for structure optimization and MD is set to 520 eV. The convergence thresholds of energy and force are set to $1\times10^{-6}$ eV and 0.001 eV/Å, respectively. The reciprocal space of brillouin zone used the Monkhorst-Pack method and the brillouin zone sampling on k-grid of spacing are set as $10\times10\times10$ for structures with surface indices of (1 0 0) and (1 1 0) and $7\times7\times7$ for structures with surface indices of (1 1 1).

Slab models containing 6 N atomic layers are used to simulate surfaces, where the atomic structures in the bottom 2 layers are fixed during structural relaxation and MD simulations. In the simulations of pressure effects on the surface stabilities, all of nitrogen and helium atoms are allowed to relax. The number of N atoms in each atomic layer is selected to 8, which is equal to the number of atoms in the unit cell of cg-N. For structural relaxation, the surface model contains 48 N atoms as shown in the Fig. 2, while 4 times larger supercells containing 192 N atoms are used to investigate the surface stabilities at finite temperature in the MD simulations with a time step of 0.5 fs (Fig. S4). To avoid image interaction, a vacuum of 20 Å is inserted between surfaces. First-principles MD simulations are run in the *NVT* ensemble for candidate structures at room temperature (300 K) and higher temperature (500 K, 750K, 1000K, and 1250

K). VAPKIT (50) and MULTIWFN (51) are also adopted in our work.

*4.2 Adsorption energies*

Adsorption energies of H on the surfaces are calculated using the following equation.

$$E_{ads} = ( E_{adsorption} - E_{pristine} - nE_i ) / n$$

Where $E_{adsorption}$, $E_{pristine}$, $E_i$, and *n* is the total energy of surfaces with H adsorption, the energy of pristine surface, the energy of the isolated H, and the number of adsorbates respectively. The $E_i$ is defined as the half energy of isolated $H_2$. Negative adsorption energy means that corresponding adsorption process is energy-favorable.

## Conflict of interest

The authors declare that they have no conflict of interest.

## Acknowledgements

This work is supported by the National Natural Science Foundation of China (NSFC) under Grant of U2030114, and CASHIPS Director's Fund. The calculations were partly performed in Center for Computational Science of CASHIPS, the ScGrid of Supercomputing Center and Computer Network Information Center of Chinese Academy of Sciences, and the Hefei Advanced Computing Center.

## Author contributions

X.W. conceived the study and planned the research. G.C and C.N. performed the calculations. W.X. and G.C. analyzed the data. X.W. and G.C wrote the manuscript. All the authors discussed the results and comment on the manuscript.

## Data availability

All data generated or analyzed during this study are included in this published article and its Supplementary Information files. The data that support the findings of this study are available from the corresponding author upon reasonable request.

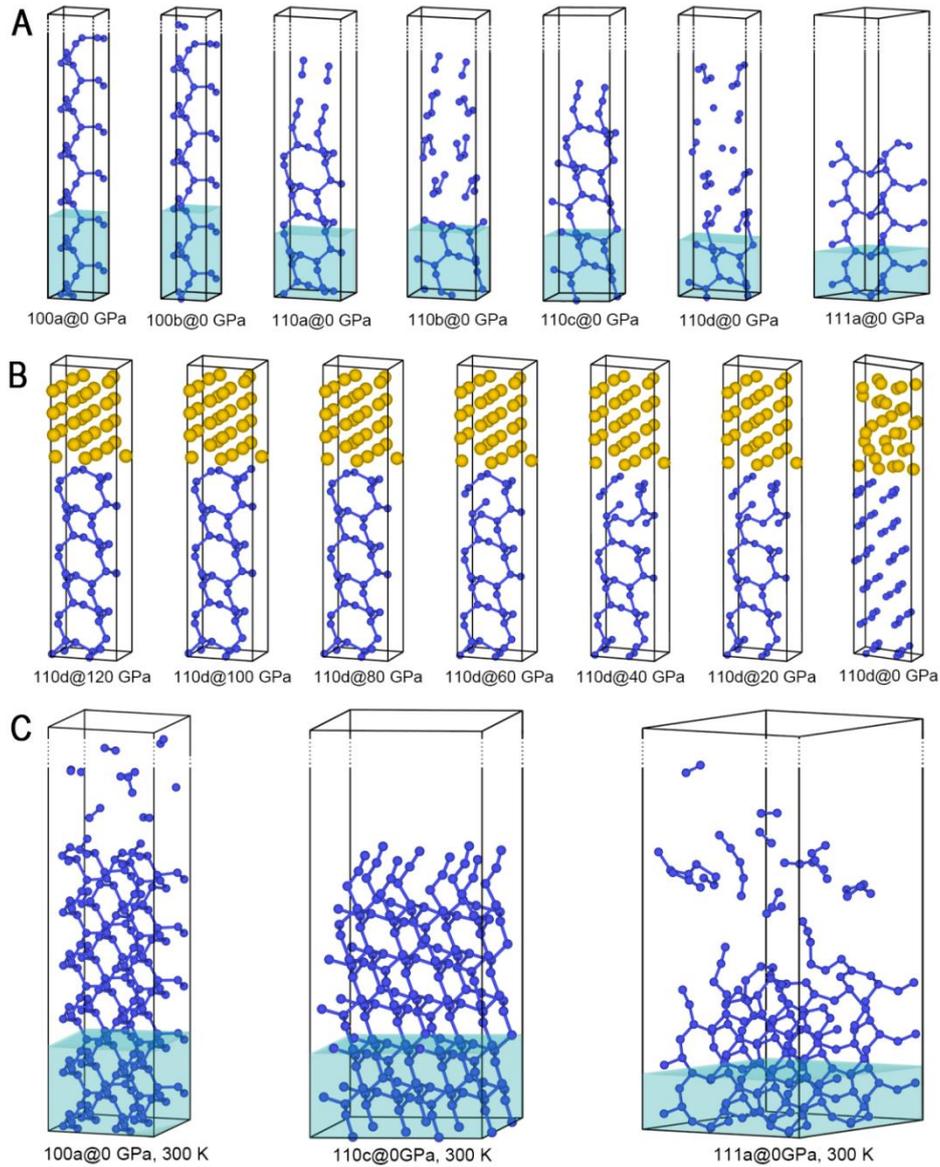

**Fig. 1.** Stabilities of pristine surfaces. (a) Structures of pristine surfaces after structural relaxations at 0 GPa. (b) Optimized structures of 110d surfaces at pressures ranging from 0 GPa to 120 GPa, and He atoms are filled in vacuum zone. (c) Structures of pristine 100a, 110c, and 111a surfaces after MD simulations at 300 K by using the supercell with 192 N atoms. The atoms located in blue shadow area are fixed. For structural relaxations filled with He at high-pressure, all nitrogen atoms and helium atoms are allowed to relax. To clearly show the atomic structure of surfaces, the vacuum between surfaces are not fully presented, and it is indicated by the dotted lines. Blue balls present nitrogen atoms and orange presents helium atoms.

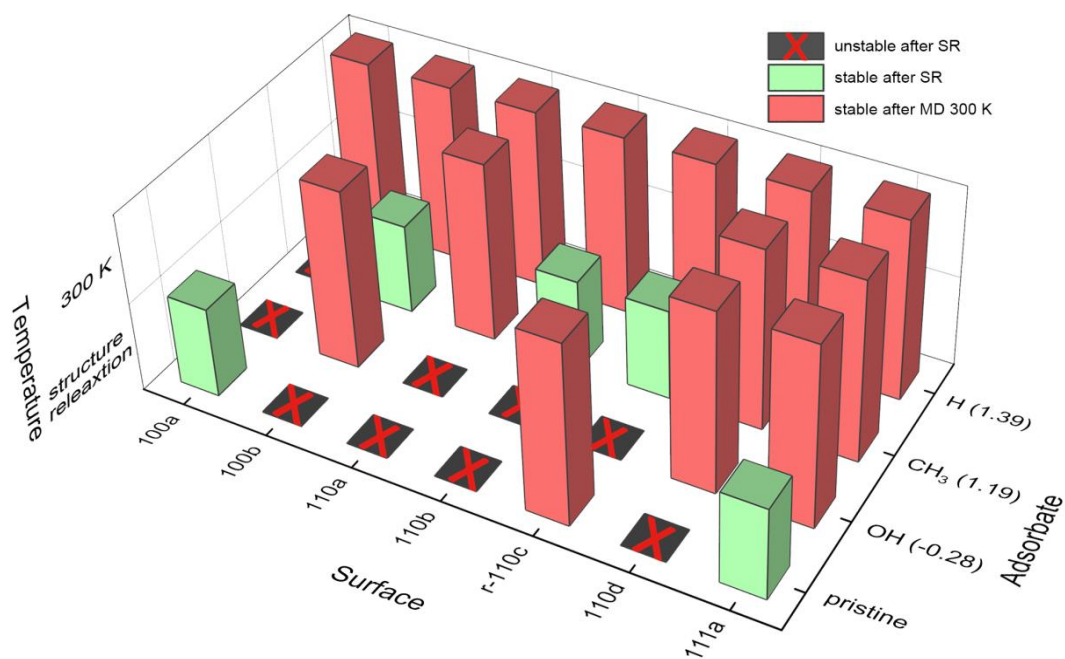

**Fig. 2.** The stabilities of surfaces without and with H, OH, and $CH_3$ saturations after structure relaxation and MD simulations at 300 K are shown. The brackets on the y-axis tags show the average number of electrons (unit: e) transferred from adsorbates to surface, where positive (negative) values present electrons transfer from adsorbates to surfaces (from surfaces to adsorbates). The r-110c means the 110c surface with surface reconfiguration.

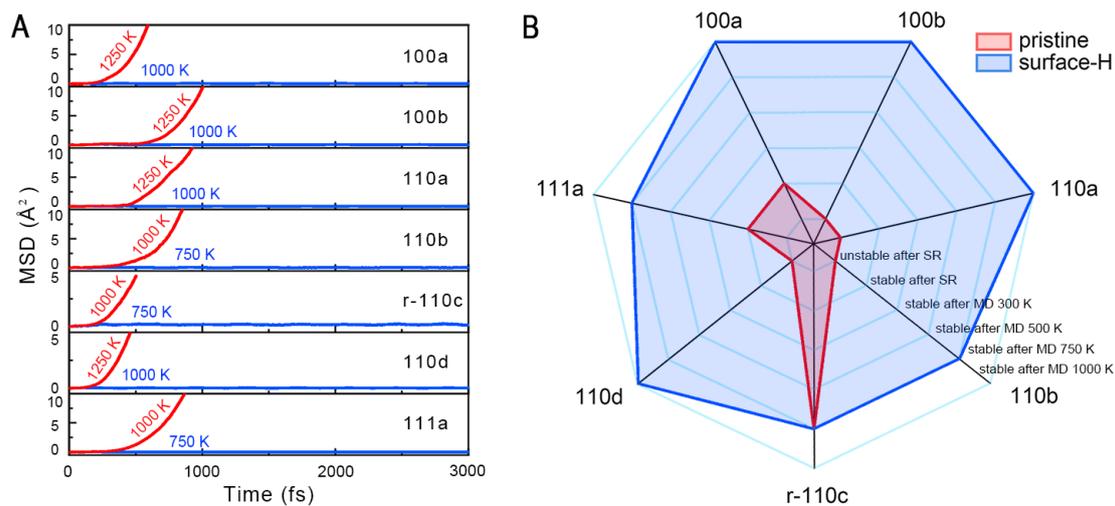

**Fig. 3.** Stabilities of surfaces with H saturation. (a) The MSD of surface-H at the temperature before (blue line) and after (red line) after structure collapse. (b) Stability of pristine surfaces and surface-H after structural relaxation and MD simulations at finite temperatures up to 1000 K. Different levels of rings represent the different stability, the more outer the ring, the more stable it is. The r-110c means the 110c surface with surface reconfiguration.

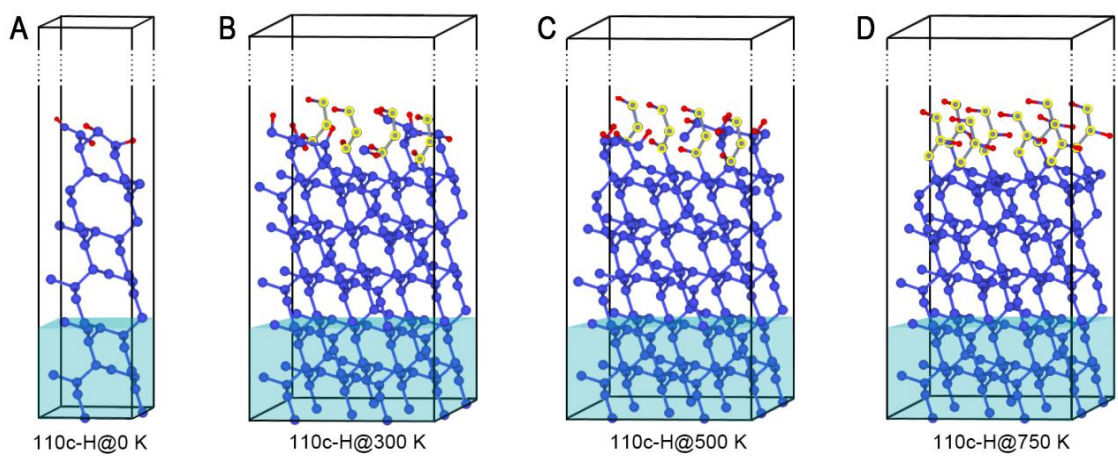

**Fig. 4.** Structures of 110c-H surfaces at different temperatures. (a) Structure of 110c-H after structure relaxation. (b) Structures of 110c-H after MD simulations at 300 K. (c) Structures of 110c-H after MD simulations at 500 K. (d) Structures of 110c-H after MD simulations at 750 K. Reconfiguration does not occur after the structural relaxation, while half of surface has reconfiguration after MD simulations at 300 K and 500 K. At 750 K, surface reconfiguration is completed. To facilitate identification, reconfigured N atoms are marked with yellow circles.

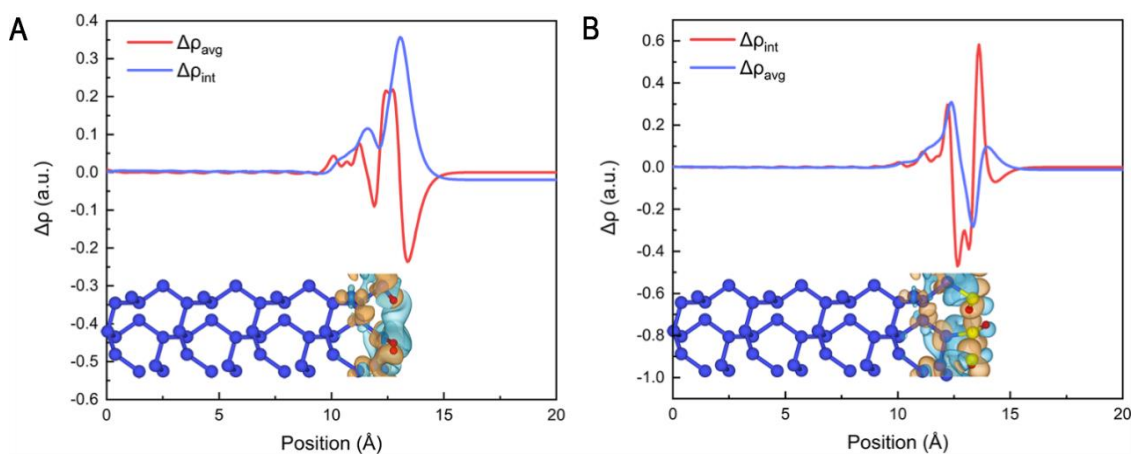

**Fig. 5.** Electron density difference (EDD). (a) The EDD of 111a-H surface. (b) The EDD of 111a-OH surface. The curves are plotted along the z direction of the structures, which are shown in the insets. Local integral curve shown in red line presents the electron gain (positive values) and loss (negative values), and integral curve shown in blue line is the integral of the local integral curve. Isosurface in the structures is set to 0.003 e/Bohr$^3$, and Orange (Blue) isosurface presents positive (negative) values of electron density difference.